\documentclass[aps,preprint,showpacs,floats,epsf,epsfig,nofootinbib,12pt]{revtex4}
\usepackage{graphicx}% Include figure files
\usepackage{epsfig}
\usepackage[dvips]{color}

\def\beq{\begin{eqnarray}}
\def\eeq{\end{eqnarray}}
\def\non{\nonumber}

\def\la{\langle}
\def\ra{\rangle}
\def\Mb{M_{b}}
\def\Mc{M_{c}}

\begin{document}

\title{ Revisiting $\Lambda_{b}\to\Lambda_{c}$ and $\Sigma_{b}\to\Sigma_{c}$ weak decays in the light-front quark model}

\vspace{1cm}

\author{ Hong-Wei Ke$^{1}$   \footnote{khw020056@hotmail.com}, Ning Hao$^{1}$ and
        Xue-Qian Li$^2$\footnote{lixq@nankai.edu.cn},
   }

\affiliation{  $^{1}$ School of Science, Tianjin University, Tianjin 300072, China
\\
  $^{2}$ School of Physics, Nankai University, Tianjin 300071, China }

\vspace{12cm}

\begin{abstract}

In this work, we study  $\Lambda_{b}\to\Lambda_{c}$ and
$\Sigma_{b}\to\Sigma_{c}$ weak decays in the light-front quark
model. As is well known, the key point for such calculations is
properly evaluating the hadronic transition matrix elements which
are dominated by the non-perturbative QCD effect. In our
calculation, we employ the light-front quark model and rather than
the traditional diquark picture, we account the two spectator
light quarks as individual ones.
Namely during the transition, they
retain their color indices, momenta and spin polarizations
unchanged. Definitely, the subsystem composed of the two light quarks is
still in a color-anti-triplet and possesses a definite spin, but
we do not priori assume the two light quarks to be in a bound
system--diquark. Our purpose is probing the diquark picture,
via comparing the results with the available data, we test
the validity and applicability of the diquark structure which
turns a three-body problem into a two-body one, so greatly
simplifies the calculation. It is indicated that the two
approaches (diquark and a subsystem within which the two light
quarks are free) lead to
similar numerical results even though the model parameters in the two schemes might
deviate slightly.
Thus, the diquark approach seems sufficiently reasonable.


\end{abstract}

\maketitle

\section{Introduction}
The study on baryon physics is much behind that on mesons so far,
because the structure of baryon is remarkably more complicated
even though only one more quark gets involved. The case of
$\Lambda_b$, $\Lambda_c$, $\Sigma_b$ and $\Sigma_c$ which contain
one heavy quark is simpler than that with all light quarks, so
that corresponding research attracts attention of both
experimentalists and theorists of high energy physics. Thanks to
the successful operation of LHC, plenty of data on baryons,
especially those on heavy baryons have been collected. Thus
researchers have a great opportunity to study heavy baryons via
their production and decays to gain information of their structure
and how the fundamental dynamics works for the baryon case.
$\Lambda_b$ is the ground state of $b$ baryons so it can only
decay via weak interactions. Indeed besides study on the baryon
structure, its weak decays may be valuable for determining the CKM
parameter $V_{cb}$ as a compensation to the measurements on mesons
and furthermore one can investigate the non-perturbative QCD
effects in the heavy baryon system because of the existence of the
heavy quark. $\Sigma_b$ is heavier than $\Lambda_b$ which would
dominantly decay via the portal $\Lambda_b+\pi$, therefore a
sizable branching ratio of its weak decays may imply a possible
involvement of new physics, so can serve as an ideal laboratory
for searching new physics beyond the standard model.

%When discussing baryon structures, there is a
%fundamental problem: if the diquark picture represents real physics?

The weak decays of heavy baryons including $\Lambda_b$ and
$\Sigma_b$ have been studied. For example: Korner and
Kroll\cite{Korner:1992uw} explored the weak decays of baryons
under the heavy quark limit\cite{IW} where the quark-diquark
picture was employed; in Ref.\cite{Ebert:2006rp} Ebert $et\, al.$
used their relativistic quark model to calculate the decay rates
of several weak decay modes where baryons consist of a heavy quark
and a light diquark; Singleton examined the semileptonic decays of
spin-$\frac{1}{2}$ baryons in the spectator-quark
model\cite{Singleton:1990ye}; in Ref.\cite{Ivanov:1996fj} Ivanov
$et\,al.$ employed their relativistic three-quark model to study
the weak decays of several baryons under the heavy quark limit;
lately, Ivanov $et\,al.$\cite{Ivanov:1998ya} also studied heavy
baryon decays in the Bethe-Salpeter approach under the heavy quark
limit. By those works the properties of the weak decays of
$\Lambda_b$ and $\Sigma_b$ have been investigated and the
non-perturbative QCD effects for baryons (at least for heavy
baryons) are partly understood or at least can be approximately
handled.

We extended the light-front quark model to study the weak decays
of $\Lambda_b$ and $\Sigma_b$ in the heavy-quark-light-diquark
picture of baryon\cite{Ke:2007tg,Wei:2009np,Ke:2012wa}. The
light-front quark model (LFQM) is a relativistic quark model which
has been applied to study transitions among mesons and the results
agree with the data within reasonable error tolerance
\cite{Jaus,Ji:1992yf,Cheng:1996if,Cheng:2003sm,Hwang:2006cua,Li:2010bb,Ke:2009ed,Wei:2009nc,Choi:2007se}.
Our results presented in \cite{Ke:2007tg,Wei:2009np,Ke:2012wa} are
consistent with those given in literatures. The application of the
extension of the light-front quark model to baryon has achieved a
preliminary success\cite{Wang:2017mqp,Chua:2018lfa,Yu:2017zst}.
Even though the baryon quark-diquark picture works well for
dealing with the transition among heavy baryons, its reasonability
and applicability are still not fully convincing yet.

There exists an acute dispute about the diquark structure of
baryons yet. As is well known, the fundamental structure of
baryons is determined by the Faddeev equations. However, that is
an equation group for the three-body system whose solutions are
difficult to gain. The diquark picture has been raised for a long
while, even at the birth time of quark
model\cite{GellMann:1964nj}. In that picture two quarks are
loosely bound into a subsystem which can be a vector, axial
vector, scalar or pseudoscalar in color-anti-triplet. This
approach definitely is an approximation which turns the three-body
problem into a two-body problem, so greatly simplifies the
calculation. In the earlier works as listed above where the
diquark picture was employed, the diquark was treated as a
point-like boson with a definite mass, spin and isospin. When it
is involved in the concerned reaction\cite{Guo:2005qa}, a form
factor composed of a few free parameters which are fixed
phenomenologically, is introduced. The picture is somehow in
analog to the case of elastic electron-proton scattering where the
inner structure of proton is manifested in the electric and
magnetic form factors.

The assertion needs to be verified in some ways.
To test the validity and applicability
of the diquark approach, in this work, we treat the three quarks (one heavy and two light)
as individual ones and possess their own color indices, spin polarizations(or helicities)
and momenta, namely they share the total momentum
of the baryon. During the transition, the two light quarks are spectators, i.e. maintain their
all quantum numbers (spin, color) and momenta unchanged. In one word, we make a three-body calculation
rather than a two-body one.
Comparing the upcoming results with that
obtained in terms of diquark, one can make a judgement whether the diquark picture indeed works well
in the concerned processes where the light-quark subsystem is a spectator. Our results show that
when the light quarks can be treated as spectators during the hadronic transitions, the
point-like diquark picture is a good approximation, at least at the leading order.

%To further investigate the phenomenological picture, we study the weak
%transitions among heavy baryons under the light-front quark model where
%we deliberately treat the three quarks in the baryon as individual subjects
%with their fixed quantum numbers(such as spin, color etc.) and momenta. Namely, we do not
%specially invoke the diquark picture to calculate the transition rate, while the two light
%quarks serve as spectators, but not as a whole subsystem. i.e. diquark to appear.

%In the traditional quark-diquark picture, the two light quarks do not
%behave as individual subjects, but are combined into a diquark, a loosely bound
%subsystem of color-anti-triplet. In this subject, no matter the diquark serves as a spectator
%during the transition, or takes part in the reaction process, it is viewed as a whole and its
%inner structure is represented by a form factor which contains a few free parameters to be fixed
%by fitting data. In this work, without invoking the diquark picture, we calculate the rate
%of weak decays of $\Lambda_b$ and then compare the results with that obtained with the diquark
%assumption. In this way, we can check the reasonability of the diquark picture, at least
%when the diqaurk (or light quarks) does not take part in the transition process as a spectator.
%In fact, it is believed that the diquark structure is a good approximation at least for the heavy
%baryons containing a heavy b or c quark.

In order to calculate the hadronic transition matrix element one
needs to know the effective vertex functions. We construct them at
first. Since the isospins of $\Lambda_b$ and $\Lambda_c$ are 0,
the light $ud$ subsystem must be an isospin-0 and color $\bar 3$
state. To guarantee the total spin of $\Lambda_{b(c)}$ to be 1/2
and the spin of the $ud$ system should be zero, i.e. the
wavefunction of the $ud$ subsystem is totally antisymmetric for
spin$\times$color$\times$isospin. Instead,  the isospin of
$\Sigma_b$ or $\Sigma_c$  is 1, according to the same principle
the spin of the light subsystem ($ud$) is determined to be 1. Thus
the spin polarizations of the two quarks are not free, but
correlated.

%Since the spin of $ud$ is a good
%quantum number, $ud$ quarks can be considered as a subsystem ( a diquark with inner degree of freedom)
%which couple to a heavy quark to form a baryon
%when the spin and spatial wave function are constructed.
With the spin arrangements \cite{Tawfiq:1998nk} we obtain the vertex function
of  $\Lambda_b$, $\Lambda_c$, $\Sigma_b$ and $\Sigma_c$ (denoting as
$\mathcal{B}_{Q(')}$).
Then following the common approach\cite{Ke:2007tg,Wei:2009np,Ke:2012wa}
we write down the transition matrix element and extract the
form factors $f_1$, $f_2$, $g_1$ and $g_2$ which are defined for the
transition(see the text for details). We compute these form factors $f_1$, $f_2$,
$g_1$ and $g_2$ numerically.

Since the leptons do not participate  in
the strong interaction, the semileptonic decay is less contaminated
by the non-perturbative QCD effect, therefore study on
semileptonic decay might more help to test the employed model and/or
constrain the model parameters. With the form factors we calculate
the widths of the concerned semileptonic decays. Comparing the numerical
result with the data the model parameters in the wave function are
fixed. Moreover, moving one more step,
using those parameters we write out the amplitude of the non-leptonic decay
$\Sigma_{b}\rightarrow \Sigma_{c}+M$. In the case, we suppose that one can factorize out
the meson current. Definitely, the factorization does not rigorously holds, but can be a good
approximation at the leading order which was thoroughly investigated for the meson decays.

This paper is organized as follows: after the introduction, in
section II we construct the vertex functions of heavy baryons,
then write down the transition amplitude for
$\Lambda_b\to\Lambda_c$ and $\Sigma_{b}\rightarrow \Sigma_{c}$ in
the light-front quark model and deduce the form factors, then we
present our numerical results for$\Lambda_b\to\Lambda_c$ and
$\Sigma_{b}\rightarrow \Sigma_{c}$ along with all necessary input
parameters in section III. Section IV is devoted to our conclusion
and discussions.

\section{ $\Lambda_{b}\to\Lambda_{c}$ and $\Sigma_{b}\to\Sigma_{c}$ in the light-front quark model}

\subsection{the vertex functions of $\Lambda_{b}$, $\Lambda_{c}$, $\Sigma_{b}$ and $\Sigma_{c}$}

In our previous work \cite{Ke:2007tg,Wei:2009np,Ke:2012wa}, we
employed the quark-diquark picture to study the transitions.
Instead, in this work we will estimate the transition rate
by treating the three quarks as individual ones. It is
noted, the transition occurs between heavy $b$ and $c$
quarks. Even though the other two light quarks are not bound together,
the subsystem where they reside in, is still of definite spin, color
and isospin and as the two quarks are spectators, all the quantum
numbers of the subsystem keep unchanged.  In analog to the
references\cite{pentaquark1,pentaquark2} the vertex functions of a
baryon $\mathcal{B}_Q$ ($Q=b,c$) with total spin $S=1/2$ and
momentum $P$ is
\begin{eqnarray}\label{eq:lfbaryon}
  && |\mathcal{B}_{Q}(P,S,S_z)\rangle=\int\{d^3\tilde p_1\}\{d^3\tilde
p_2\}\{d^3\tilde p_3\} \,
  2(2\pi)^3\delta^3(\tilde{P}-\tilde{p_1}-\tilde{p_2}-\tilde{p_3}) \nonumber\\
 &&\times\sum_{\lambda_1,\lambda_2,\lambda_3}\Psi^{SS_z}(\tilde{p}_1,\tilde{p}_2,\tilde{p}_3,\lambda_1,\lambda_2,\lambda_3)
  C^{\alpha\beta\gamma}F_{Qdu}\left|\right.
  Q_{\alpha}(p_1,\lambda_1)u_{\beta}(p_2,\lambda_2)d_{\gamma}(p_3,\lambda_3)\ra.
\end{eqnarray}
where  $C^{\alpha\beta\gamma}$ and $F_{Qdu}$ are the color and
flavor factors, $\lambda_i\,(i=1,2,3)$ and $p_i\,(i=1,2,3)$ are
helicities and light-front momenta of the on-mass-shell quarks
defined as
\begin{equation}
 \tilde{p_i}=(p_i^+,p_{i\perp}),\qquad p_{i\perp}=(p_i^1,p_i^2),\qquad
 p_i^-=\frac{m^2+p_{i\perp}^2}{p_i^+},\qquad\{d^3p_i\}\equiv\frac{dp_i^+d^2
 p_{i\perp}}{2(2\pi)^3}.
\end{equation}

In order to describe the motions of the constituents, one
needs to introduce intrinsic variables $(x_i, k_{i\perp})$ (
$i=1,2$) through
\begin{eqnarray}
&&p^+_i=x_i P^+, \qquad p_{i\perp}=x_i P_{\perp}+k_{i\perp}
 \qquad x_1+x_2+x_3=1, \qquad k_{1\perp}+k_{2\perp}+k_{3\perp}=0,
\end{eqnarray}
where $x_i$ are the light-front momentum fractions constrained by
$0<x_1, x_2, x_3<1$. The variables $(x_i, k_{i\perp})$ are
independent of the total momentum of the hadron and thus are
Lorentz-invariant. The invariant mass square $M_0^2$ is
defined as
 \begin{eqnarray} \label{eq:Mpz}
  M_0^2=\frac{k_{1\perp}^2+m_1^2}{x_1}+
        \frac{k_{2\perp}^2+m_2^2}{x_2}+\frac{k_{3\perp}^2+m_3^2}{x_3}.
 \end{eqnarray}
The invariant mass $M_0$ is in general different from the hadron
mass $M$ which obeys the physical mass-shell condition
$M^2=P^2$. This is due to the fact that the baryon, heavy quark
and the two-light-quark subsystem cannot be on their mass shells simultaneously. We
define the internal momenta as
 \beq
 k_i=(k_i^-,k_i^+,k_{i\bot})=(e_i-k_{iz},e_i+k_{iz},k_{i\bot})=
  (\frac{m_i^2+k_{i\bot}^2}{x_iM_0},x_iM_0,k_{i\bot}).
 \eeq
It is easy to obtain
 \begin{eqnarray}
  e_i&=&\frac{x_iM_0}{2}+\frac{m_i^2+k_{i\perp}^2}{2x_iM_0}
 ,\non\\
 k_{iz}&=&\frac{x_iM_0}{2}-\frac{m_i^2+k_{i\perp}^2}{2x_iM_0},
 \end{eqnarray}
where $e_i$ denotes the energy of the $i$-th constituent. The
momenta $k_{i\bot}$ and $k_{iz}$ constitute a momentum vector
$\vec k_i=(k_{i\bot}, k_{iz})$ and correspond to the components in
the transverse and $z$ directions, respectively.

%In many works\cite{Korner:1992uw,Ebert:2006rp}  heavy baryon is regarded as in the
%heavy-quark-light-diquark structure where the diquark is a point-like particle.
%The light diquark is in a color $\bar 3$ state, thus the color
%indices of the two quarks are antisymmetric. As a consequence, the
%other quantum states of the diquark wave-function must be totally
%symmetric (flavor-spin-orbit) due to the Pauli-principle. We
%suppose the relative orbital angular momentum between the two
%quarks is zero, i.e. the diquark is in the $S$-state, thus the
%flavor-spin part should be symmetric. Since the isospin of
%$\Sigma_b$ or $\Sigma_c$ is 1, the diquark containing $ud$ flavors
%is in an $I=1$ state, namely its flavor part is symmetric. It
%demands its spin part to be symmetric either, i.e. the diquark is
%an axial vector. By contrary, for $\Lambda_c$ or $\Lambda_b$,
%whose isospin is 0, the $ud$ must be an isospin-0 state, i.e.
%flavor-anti-symmetric state. Thus the spin of the diquark must be
%0, i.e. the diquark is a scalar. In this paper, we will employ
%three-quark picture instead of the quark-diquark picture for
%baryon. However the total spin of light $ud$ system is good
%quantum number and we can take the light $ud$ system for a
%subsystem (or a diquark with inner degree of freedom) which
%couples to the heavy quark to form a heavy baryon.

Being enlightened by \cite{Tawfiq:1998nk} the spin and spatial
wave function for  $\Lambda_Q$ is written as
\begin{eqnarray}
\Psi^{SS_z}_0(\tilde{p}_i,\lambda_i)=&&A_0 \bar
U(p_3,\lambda_3)[(\bar
P\!\!\!\!\slash+M_0)\gamma_5]V(p_2,\lambda_2)\bar
U_Q(p_1,\lambda_1) U(\bar P,S) \varphi(x_i,k_{i\perp}),
\end{eqnarray}
and  for $\Sigma_Q$
\begin{eqnarray}
\Psi^{SS_z}_1(\tilde{p}_i,\lambda_i)=&&A_1 \bar
U(p_3,\lambda_3)[(\bar
P\!\!\!\!\slash+M_0)\gamma_{\perp\alpha}]V(p_2,\lambda_2)\bar
U_Q(p_1,\lambda_1) \gamma_{\perp\alpha}\gamma_{5}U(\bar
P,S)\varphi(x_i,k_{i\perp}),
\end{eqnarray}
where $\Psi^{SS_z}_0(\tilde{p}_i,\lambda_i)$ represents
$\Psi^{SS_z}(\tilde{p}_1,\tilde{p}_2,\tilde{p}_3,\lambda_1,\lambda_2,\lambda_3)$,
$U,\,V$ and $\bar U$ are spinors of the quarks,
$\varphi(x_i,k_{i\perp})$ denotes
$\varphi(x_1,x_2,x_3,k_{1\perp},k_{2\perp},k_{3\perp})$, $p_1$ is
the the momentum of the heavy quark $Q$, $\,p_2\,,p_3$ are the
momenta of the two light quarks, $\bar P=p_1+p_2+p_3$, $
\gamma_{\perp\alpha}=\gamma_{\alpha}-v\!\!\!\slash v_{\alpha}$,and
 $\lambda_1,\lambda_2, \lambda_3$ are the helicities of the constituents.

With the normalization of the state $|\mathcal{B}_{Q}\rangle$
\beq\label{A121}
 \la
 \mathcal{B}_{Q}(P',S',S'_z)|\mathcal{B}_{Q}(P,S,S_z)\ra=2(2\pi)^3P^+
  \delta^3(\tilde{P}'-\tilde{P})\delta_{S'S}\delta_{S'_zS_z},
 \eeq
and \beq\label{A122}
\int(\prod^3_{i=1}\frac{dx_id^2k_{i\perp}}{2(2\pi)^3})2(2\pi)^3\delta(1-\sum
x_i)\delta^2(\sum
k_{i\perp})\varphi^*(x_i,k_{i\perp})\varphi(x_i,k_{i\perp})=1,
 \eeq
one can obtain
\begin{eqnarray}
A_0%&&=\frac{1}{4\sqrt{P^+(M_0m_1+p_1\cdot\bar{P})(m_2m_3M_0^2+m_3M_0p_2\cdot\bar{P}+
%m_2M_0p_3\cdot\bar{P}+p_2\cdot\bar{P}p_3\cdot\bar{P})}}\nonumber\\
&&=\frac{1}{4\sqrt{P^+M_0^3(m_1+e_1)(m_2+e_2)(m_3+e_3)}},\nonumber\\
A_1
%&&=\frac{1}{4\sqrt{3P^+(M_0m_1+p_1\cdot\bar{P})(M_0m_2+p_2\cdot\bar{P})(M_0m_3+p_3\cdot\bar{P})}}\nonumber\\
&&=\frac{1}{4\sqrt{3P^+M_0^3(m_1+e_1)(m_2+e_2)(m_3+e_3)}},
\end{eqnarray}
where $p_i\cdot\bar{P}=e_iM_0\,(i=1,2,3)$ is used.

The spatial wave function is\cite{pentaquark1,pentaquark2}
 \beq\label{A122}
\varphi(x_1,x_2,x_3,k_{1\perp},k_{2\perp},k_{3\perp})=\frac{e_1e_2e_3}{x_1x_2x_3M_0}
\varphi(\overrightarrow{k}_1,\beta_1)\varphi(\frac{\overrightarrow{k}_2-\overrightarrow{k}_3}{2},\beta_{23})
 \eeq
with
$\varphi(\overrightarrow{k},\beta)=4(\frac{\pi}{\beta^2})^{3/4}{\rm
exp}(\frac{-k_z^2-k^2_\perp}{2\beta^2})$.
% which can be obtained by normalizing
%the state  $|\Sigma_Q(P,S,S_z)\rangle$ , \beq\label{A12}
% \la
% \Sigma_Q(P',S',S'_z)|\Sigma_Q(P,S,S_z)\ra=2(2\pi)^3P^+
%  \delta^3(\tilde{P}'-\tilde{P})\delta_{S'S}\delta_{S'_zS_z}.
% \eeq

\subsection{the  form factors of $\Lambda_{b}\to\Lambda_{c}$ in LFQM}

\begin{figure}
\begin{center}
%\begin{tabular}{ccc}
\scalebox{0.8}{\includegraphics{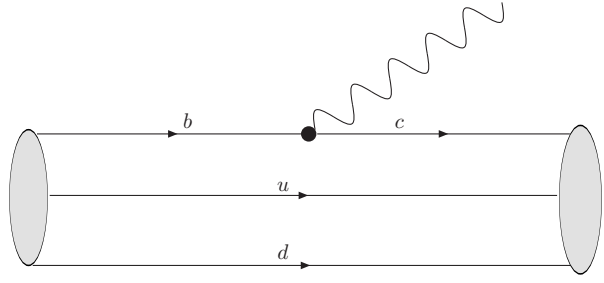}}
%\end{tabular}
\end{center}
\caption{The Feynman diagram for $\Lambda_{b}\to\Lambda_{c}$ and
$\Sigma_{b}\to\Sigma_{c}$ transitions, where $\bullet$ denotes
$V-A$ current vertex.}\label{t1}
\end{figure}

The form factors for the weak transition
$\Lambda_{b}\to\Lambda_{c}$ are defined in the standard way as
\begin{eqnarray}\label{s1}
&& \la \Lambda_{c}(P',S',S_z') \mid \bar{c}\gamma_{\mu}
 (1-\gamma_{5})b\mid \Lambda_{b}(P,S,S_z) \ra  \non \\
 &=& \bar{U}_{\Lambda_{c}}(P',S'_z) \left[ \gamma_{\mu} f^s_{1}(q^{2})
 +i\sigma_{\mu \nu} \frac{ q^{\nu}}{M_{\Lambda_{b}}}f^s_{2}(q^{2})
 +\frac{q_{\mu}}{M_{\Lambda_{b}}} f^s_{3}(q^{2})
 \right] U_{\Lambda_{b}}(P,S_z) \nonumber \\
 &&-\bar U_{\Lambda_{c}}(P',S'_z)\left[\gamma_{\mu} g^s_{1}(q^{2})
  +i\sigma_{\mu \nu} \frac{ q^{\nu}}{M_{\Lambda_{b}}}g^s_{2}(q^{2})+
  \frac{q_{\mu}}{M_{\Lambda_{b}}}g^s_{3}(q^{2})
 \right]\gamma_{5} U_{\Lambda_{b}}(P,S_z).
\end{eqnarray}
Since $S=S'=1/2$, we write
$\mid\Lambda_{b}(P,S,S_z)\ra$ and $\mid\Lambda_{c}(P',S',S'_z)\ra$
as $\mid\Lambda_{b}(P,S_z)\ra$ and $\mid\Lambda_{c}(P',S'_z)\ra$
respectively.

The lowest order Feynman diagram responsible for the
$\Lambda_{b}\to\Lambda_{c}$ weak decay is shown in Fig. \ref{t1}.
Following the approach given in Refs.
\cite{pentaquark1,pentaquark2,Ke:2007tg,Ke:2012wa} the transition
matrix element can be calculated with the vertex functions of
$\mid \Lambda_{b}(P,S_z) \ra$ and $\mid \Lambda_{c}^(P',,S'_z)
\ra$ supposing $u$ and $d$ quarks are spectators,

\begin{eqnarray}\label{s2}
&& \la \Lambda_{c}(P',S_z') \mid \bar{c}
\gamma^{\mu} (1-\gamma_{5}) b \mid \Lambda_{b}(P,S_z) \ra  \nonumber \\
 &=& \int\frac{\{d^3 \tilde p_2\}\{d^3 \tilde p_3\}\phi_{\Lambda_{c}}^*(x',k'_{\perp})
  \phi_{\Lambda_{b}}(x,k_{\perp}){\rm Tr}[(\bar{P'}\!\!\!\!\!\slash'-M_0')\gamma_{5}(p_2\!\!\!\!\!\slash+m_2)
  (\bar{P}\!\!\!\!\!\slash'+M_0)\gamma_{5}(p_3\!\!\!\!\!\slash-m_3)]}{16\sqrt{p^+_1p'^+_1{P}^+{P'}^+M_0^3(m_1+e_1)
 (m_2+e_2)(m_3+e_3)(m_1'+e_1')
 (m_2'+e_2')(m_3'+e_3')}}\nonumber \\
  &&\times  \bar{U}(\bar{P'},S'_z)
  (p_1\!\!\!\!\!\slash'+m'_1)\gamma^{\mu}(1-\gamma_{5})
  (p_1\!\!\!\!\!\slash+m_1)  U(\bar{P},S_z),
\end{eqnarray}
where
 \beq
m_1=m_b, \qquad m'_1=m_c, \qquad m_2=m_{u}, \qquad m_3=m_{d}
 \eeq
and $Q$ ($Q'$) represents the heavy quark $b$ ($c$), $p_1$
($p'_1$) denotes the four-momentum of the heavy quark $b$ ($c$),
$P$ ($P'$) stands as the four-momentum of $\Lambda_{b}$
($\Lambda_{c}$). Setting $\tilde{p}_2=\tilde{p}'_2$, we have
 \beq
 x'=\frac{P^+}{P^{'+}}x, \qquad \qquad
 k'_{\perp}=k_{\perp}+x_2q_{\perp}.
 \eeq

In terms of the approach given in Ref.\cite{pentaquark1} we extract the
form factors defined in Eq. (\ref{s1}) from the Eq. (\ref{s2})
\begin{eqnarray}\label{s21}
f^s_1
 &=& \int\frac{ d x_2 d^2 k_{2\perp}}{2(2\pi)^3}\frac{ d x_3 d^2 k_{3\perp}}{2(2\pi)^3}
 \frac{{\rm Tr}[(\bar{P'}\!\!\!\!\!\slash-M_0')\gamma_{5}(p_2\!\!\!\!\!\slash+m_2)(\bar{P}\!\!\!\!\!\slash+M_0)
 \gamma_{5}(p_3\!\!\!\!\!\slash-m_3)]}{\sqrt{M_0^3(m_1+e_1)
 (m_2+e_2)(m_3+e_3)(m_1'+e_1')
 (m_2'+e_2')(m_3'+e_3')}}\nonumber \\
  &&\times\frac{\phi_{\Lambda_{c}}^*(x',k'_{\perp})
  \phi_{\Lambda_{b}}(x,k_{\perp})}{16\sqrt{x_1x'_1}} \frac{{\rm Tr}[ (\bar{P}\!\!\!\!\!\slash+M_0)\gamma^+
  (\bar{P'}\!\!\!\!\!\slash+M_0')
  (p_1\!\!\!\!\!\slash'+m'_1)\gamma^{+}
  (p_1\!\!\!\!\!\slash+m_1)  ]}{8P^+P'^+},\nonumber\\
\frac{f^s_2}{M_{\Lambda_{b}}}
 &=& \frac{-i}{q_{\perp}^i}\int\frac{ d x_2 d^2 k_{2\perp}}{2(2\pi)^3}\frac{ d x_3 d^2 k_{3\perp}}{2(2\pi)^3}
 \frac{{\rm Tr}[(\bar{P'}\!\!\!\!\!\slash-M_0')\gamma_{5}(p_2\!\!\!\!\!\slash+m_2)(\bar{P}\!\!\!\!\!\slash+M_0)
 \gamma_{5}(p_3\!\!\!\!\!\slash-m_3)]}{\sqrt{M_0^3(m_1+e_1)
 (m_2+e_2)(m_3+e_3)(m_1'+e_1')
 (m_2'+e_2')(m_3'+e_3')}}\nonumber \\
  &&\times\frac{\phi_{\Lambda_{c}}^*(x',k'_{\perp})
  \phi_{\Lambda_{b}}(x,k_{\perp})}{16\sqrt{x_1x'_1}} \frac{{\rm Tr}[ (\bar{P}\!\!\!\!\!\slash+M_0)\sigma^{i+}
  (\bar{P'}\!\!\!\!\!\slash+M_0')
  (p_1\!\!\!\!\!\slash'+m'_1)\gamma^+
  (p_1\!\!\!\!\!\slash+m_1)  ]}{8P^+P'^+},
\nonumber\\
g^s_1
 &=& \int\frac{ d x_2 d^2 k_{2\perp}}{2(2\pi)^3}\frac{ d x_3 d^2 k_{3\perp}}{2(2\pi)^3}
 \frac{{\rm Tr}[(\bar{P'}\!\!\!\!\!\slash-M_0')\gamma_{5}(p_2\!\!\!\!\!\slash+m_2)
 (\bar{P}\!\!\!\!\!\slash+M_0)\gamma_{5}(p_3\!\!\!\!\!\slash-m_3)]}{\sqrt{M_0^3(m_1+e_1)
 (m_2+e_2)(m_3+e_3)(m_1'+e_1')
 (m_2'+e_2')(m_3'+e_3')}}\nonumber \\
  &&\times\frac{\phi_{\Lambda_{c}}^*(x',k'_{\perp})
  \phi_{\Lambda_{b}}(x,k_{\perp})}{16\sqrt{x_1x'_1}} \frac{{\rm Tr}
  [ (\bar{P}\!\!\!\!\!\slash+M_0)\gamma^+\gamma_5(\bar{P'}\!\!\!\!\!\slash+M_0')
  (p_1\!\!\!\!\!\slash'+m'_1)\gamma^{+}\gamma_5
  (p_1\!\!\!\!\!\slash+m_1)  ]}{8P^+P'^+},
\nonumber\\
\frac{g^s_2}{M_{\Lambda_{b}}}
 &=& \frac{i}{q_{\perp}^i}\int\frac{ d x_2 d^2 k_{2\perp}}{2(2\pi)^3}\frac{ d x_3 d^2 k_{3\perp}}{2(2\pi)^3}
 \frac{{\rm Tr}[(\bar{P'}\!\!\!\!\!\slash-M_0')\gamma_{5}(p_2\!\!\!\!\!\slash+m_2)
 (\bar{P}\!\!\!\!\!\slash+M_0)\gamma_{5}(p_3\!\!\!\!\!\slash-m_3)]}{\sqrt{M_0^3(m_1+e_1)
 (m_2+e_2)(m_3+e_3)(m_1'+e_1')
 (m_2'+e_2')(m_3'+e_3')}}\nonumber \\
  &&\times\frac{\phi_{\Lambda_{c}}^*(x',k'_{\perp})
  \phi_{\Lambda_{b}}(x,k_{\perp})}{16\sqrt{x_1x'_1}} \frac{{\rm Tr}[
  (\bar{P}\!\!\!\!\!\slash+M_0)\sigma^{i+}\gamma_5(\bar{P'}\!\!\!\!\!\slash+M_0')
  (p_1\!\!\!\!\!\slash'+m'_1)\gamma^+\gamma_5
  (p_1\!\!\!\!\!\slash+m_1)  ]}{8P^+P'^+}.
\end{eqnarray}
Expanding the traces in above formulas is straightforward, but the expressions are rather tedious,
thus we collect he explicit expressions in Appendix A. It
is noted that the form factors $f_3(q^2)$ and $g_3(q^2)$ cannot be
extracted in terms of the above method for we have imposed the
condition $q^+=0$. However, they do not contribute to the
semi-leptonic decays $\Lambda_b\to\Lambda_c l\bar\nu_l$ if
 the masses of electron and neutrino are ignored.

\subsection{The  form factors of $\Sigma_{b}\to\Sigma_{c}$ in LFQM}

Similarly, the hadronic matrix element for transition $\Sigma_{b}\to\Sigma_c$ can also be
obtained with the vertex functions of $\mid \Sigma_{b}(P,S_z) \ra$
and $\mid \Sigma_{c}^(P',,S'_z) \ra$,
\begin{eqnarray}\label{s3}
&& \la \Sigma_{c}(P',S_z') \mid \bar{c}
\gamma^{\mu} (1-\gamma_{5}) b \mid \Sigma_{b}(P,S_z) \ra  \nonumber \\
 &=& \int\{d^3 \tilde p_2\}\{d^3 \tilde p_3\}\frac{\phi_{\Sigma_{c}}^*(x',k'_{\perp})
  \phi_{\Sigma_{b}}(x,k_{\perp}){\rm Tr}[\gamma_{\perp}^\alpha(\bar{P'}\!\!\!\!\!\slash'+M_0')
  \gamma_{5}(p_2\!\!\!\!\!\slash+m_2)
  (\bar{P}\!\!\!\!\!\slash'+M_0)\gamma_{5}\gamma_{\perp}^\beta(p_3\!\!\!\!\!\slash-m_3)]}
  {48\sqrt{p^+_1p'^+_1\bar{P}^+\bar{P'}^+M_0^3(m_1+e_1)
 (m_2+e_2)(m_3+e_3)(m_1'+e_1')
 (m_2'+e_2')(m_3'+e_3')}}\nonumber \\
  &&\times  \bar{U}(\bar{P'},S'_z)\gamma_{\perp\alpha}\gamma_{5}
  (p_1\!\!\!\!\!\slash'+m'_1)\gamma^{\mu}(1-\gamma_{5})
  (p_1\!\!\!\!\!\slash+m_1)
  \gamma_{\perp\beta}\gamma_{5}U(\bar{P},S_z).
\end{eqnarray}

For the transition some form factors can also be defined  as in
Eq.(\ref{s1}), while $f^v_i$ and $g^v_i$ replace $f^s_i$
and $g^s_i$ ($i=1,2,3$) in Eq.(\ref{s1}).

The expressions of the form factors are
\begin{eqnarray}\label{s31}
f^v_1
 &=&\int\frac{ d x_2 d^2 k_{2\perp}}{2(2\pi)^3}\frac{ d x_3 d^2 k_{3\perp}}{2(2\pi)^3}
 \frac{{\rm Tr}[\gamma_{\perp}^\alpha(\bar{P'}\!\!\!\!\!\slash'+M_0')\gamma_{5}
(p_2\!\!\!\!\!\slash+m_2)(\bar{P}\!\!\!\!\!\slash'+M_0)\gamma_{5}\gamma_{\perp}^\beta(p_3\!\!\!\!\!\slash-m_3)]]}{\sqrt{M_0^3(m_1+e_1)
(m_2+e_2)(m_3+e_3)(m_1'+e_1')
(m_2'+e_2')(m_3'+e_3')}}\nonumber \\
&&\times\frac{\phi_{\Sigma_{c}}^*(x',k'_{\perp})
\phi_{\Sigma_{b}}(x,k_{\perp})}{48\sqrt{x_1x'_1}}  \frac{{\rm Tr}[ (\bar{P}\!\!\!\!\!\slash+M_0)\gamma^+(\bar{P'}\!\!\!\!\!\slash+M_0')
\gamma_{\perp\alpha}\gamma_{5}(p_1\!\!\!\!\!\slash'+m'_1)\gamma^{+}
(p_1\!\!\!\!\!\slash+m_1)\gamma_{\perp\beta}\gamma_{5}  ]}{8P^+P'^+},
\nonumber\\
\frac{f^v_2}{M_{\Sigma_{b}}} &=& \frac{-i}{q_{\perp}^i}\int\frac{
d x_2 d^2 k_{2\perp}}{2(2\pi)^3}\frac{ d x_3 d^2
k_{3\perp}}{2(2\pi)^3} \frac{{\rm
Tr}[\gamma_{\perp}^\alpha(\bar{P'}\!\!\!\!\!\slash'+M_0')\gamma_{5}(p_2\!\!\!\!\!\slash+m_2)
(\bar{P}\!\!\!\!\!\slash'+M_0)\gamma_{5}\gamma_{\perp}^\beta(p_3\!\!\!\!\!\slash-m_3)]}{\sqrt{M_0^3(m_1+e_1)
(m_2+e_2)(m_3+e_3)(m_1'+e_1')
(m_2'+e_2')(m_3'+e_3')}}\nonumber \\
&&\times\frac{\phi_{\Sigma_{c}}^*(x',k'_{\perp})
\phi_{\Sigma_{b}}(x,k_{\perp})}{48\sqrt{x_1x'_1}} \frac{{\rm Tr}[ (\bar{P}\!\!\!\!\!\slash-M_0)\sigma^{i+}(\bar{P'}\!\!\!\!\!\slash-M_0')
\gamma_{\perp\alpha}\gamma_{5}(p_1\!\!\!\!\!\slash'+m'_1)\gamma^{+}
(p_1\!\!\!\!\!\slash+m_1)\gamma_{\perp\beta}\gamma_{5}   ]}{8P^+P'^+},
\nonumber\\
g^v_1
 &=& \int\frac{ d x_2 d^2 k_{2\perp}}{2(2\pi)^3}\frac{ d x_3 d^2 k_{3\perp}}{2(2\pi)^3}
 \frac{{\rm Tr}[\gamma_{\perp}^\alpha(\bar{P'}\!\!\!\!\!\slash'+M_0')\gamma_{5}(p_2\!\!\!\!\!\slash+m_2)
(\bar{P}\!\!\!\!\!\slash'+M_0)\gamma_{5}\gamma_{\perp}^\beta(p_3\!\!\!\!\!\slash-m_3)]}{\sqrt{M_0^3(m_1+e_1)
 (m_2+e_2)(m_3+e_3)(m_1'+e_1')
 (m_2'+e_2')(m_3'+e_3')}}\nonumber \\
  &&\times\frac{\phi_{\Sigma_{c}}^*(x',k'_{\perp})
  \phi_{\Sigma_{b}}(x,k_{\perp})}{48\sqrt{x_1x'_1}} \frac{{\rm Tr}[
(\bar{P}\!\!\!\!\!\slash-M_0)\gamma^+\gamma_5(\bar{P'}\!\!\!\!\!\slash-M_0')
  \gamma_{\perp\alpha}\gamma_{5}(p_1\!\!\!\!\!\slash'+m'_1)\gamma^{+}
  (p_1\!\!\!\!\!\slash+m_1)\gamma_{\perp\beta}\gamma_{5}  ]}{8P^+P'^+},
\nonumber\\
\frac{g^v_2}{M_{\Sigma_{b}}}
 &=& \frac{i}{q_{\perp}^i}\int\frac{ d x_2 d^2 k_{2\perp}}{2(2\pi)^3}\frac{ d x_3 d^2 k_{3\perp}}{2(2\pi)^3}
 \frac{{\rm Tr}[\gamma_{\perp}^\alpha(\bar{P'}\!\!\!\!\!\slash'+M_0')\gamma_{5}(p_2\!\!\!\!\!\slash+m_2)
(\bar{P}\!\!\!\!\!\slash'+M_0)\gamma_{5}\gamma_{\perp}^\beta(p_3\!\!\!\!\!\slash-m_3)]}{\sqrt{M_0^3(m_1+e_1)
 (m_2+e_2)(m_3+e_3)(m_1'+e_1')
 (m_2'+e_2')(m_3'+e_3')}}\nonumber \\
  &&\times\frac{\phi_{\Sigma_{c}}^*(x',k'_{\perp})
  \phi_{\Sigma_{b}}(x,k_{\perp})}{48\sqrt{x_1x'_1}} \frac{{\rm Tr}[
(\bar{P}\!\!\!\!\!\slash-M_0)\sigma^{i+}\gamma_5(\bar{P'}\!\!\!\!\!\slash-M_0')
  \gamma_{\perp\alpha}\gamma_{5}(p_1\!\!\!\!\!\slash'+m'_1)\gamma^{+}
  (p_1\!\!\!\!\!\slash+m_1)\gamma_{\perp\beta}\gamma_{5}
  ]}{8P^+P'^+}.
\end{eqnarray}

The interested readers can refer to  appendix A to simplify the
form factors before  numerically  evaluating.

\section{Numerical Results}

\begin{table}
\caption{Quark mass and the parameter $\beta$ (in units of
 GeV).}\label{Tab:t1}
\begin{ruledtabular}
\begin{tabular}{cccc}
 $m_b$& $m_c$  & $m_s$  &$m_{u}$ \\\hline
 4.64& $1.3$  & $0.37$  & 0.26
\end{tabular}
\end{ruledtabular}
\end{table}

\subsection{The $\Lambda_{b}\to \Lambda_c$  form factors and some decay modes }
In order to evaluate these form factors numerically one needs the
parameters of the concerned model. Here we employ the masses of
quarks presented in Ref.\cite{Cheng:2003sm} and list them in table
I. Indeed, we know very little about the parameters $\beta_1$ and
$\beta_{23}$ in the wave function of the initial baryon and
$\beta_1'$ and $\beta_{23}'$ in that of the final baryon.
Generally the reciprocal of $\beta$ is related to the electrical
radium of the baryon. Since the strong interaction between $q$ and
$q'$ is half of that between $q\bar q'$ if it is a Coulomb-like
potential one can expect the the electrical radium of $qq'$ to be
$1/\sqrt{2}$ times that of $q\bar q'$ i.e.
$\beta_{qq'}\approx\sqrt{2}\beta_{q\bar q'}$.   In
Ref.\cite{LeYaouanc:1988fx} in terms of the binding energy the
authors also obtained the same results, so in our work we use
these $\beta$'s values which were obtained in the mesons case
\cite{Chang:2018zjq}. With these
parameters we calculate the form factors and make theoretical predictions on the
transition rates. We set $\beta_1\approx \sqrt{2}\beta_{b\bar s}$,
$\beta_1'\approx \sqrt{2}\beta_{c\bar s}$ and suppose
$\beta_2=\beta_2'\approx \sqrt{2} \beta_{u\bar d}$. It is found that
the predicted width of the semilepton decay is larger than the data. Then we
readjust the parameter $\beta_2$ to reduce the
theoretical prediction
to be closer to the data and thus we could fix the parameter. At last we
obtain $\beta_1=0.851$ GeV, $\beta'_1=0.760$ GeV and
$\beta_2=\beta_2'=0.911$ GeV. Here $\beta_2$ and $\beta_2'$ are
close to $2.9 \beta_{u\bar d}$ , which means the distance between
$u$ and $d$ quarks is smaller than the normal situation where $u$
and $d$ are evenly distributed in the inner space of the baryon.
Indeed, it implies the diquark structure.

It is noted, we derive the form factors in terms of the LFQM in
the space-like region, thus applying them to evaluate the
transition rates,  one needs to extend them to the physical region
(time-like). Following the standard scheme, we derive the form
factors for the baryonic transitions.  The procedure for the
derivation was depicted in literature and our previous work in all
details, so that we do not keep it in the context, but for
readers' convenience, relevant stuff is retained in the attached
appendix. Then, we explicitly list the corresponding parameters
which are needed for numerical computations in table
\ref{Tab:t21}.

\begin{table}
\caption{The $\Lambda_{b}\to \Lambda_c$ form factors given in the
  three-parameter form.}\label{Tab:t21}
\begin{ruledtabular}
\begin{tabular}{cccc}
  $F$    &  $F(0)$ &  $a$  &  $b$ \\\hline
  $f^s_1$  &   0.488    &  1.04    & 0.38   \\
$f^s_2$  &   -0.180    &   1.71    & 0.58   \\
 $g^s_1$  &      0.470   &     0.953 &  0.361  \\
  $g^s_2$  &      -0.0479   &    2.06  &  0.89
\end{tabular}
\end{ruledtabular}
\end{table}

\begin{figure}[hhh]
\begin{center}
\scalebox{0.8}{\includegraphics{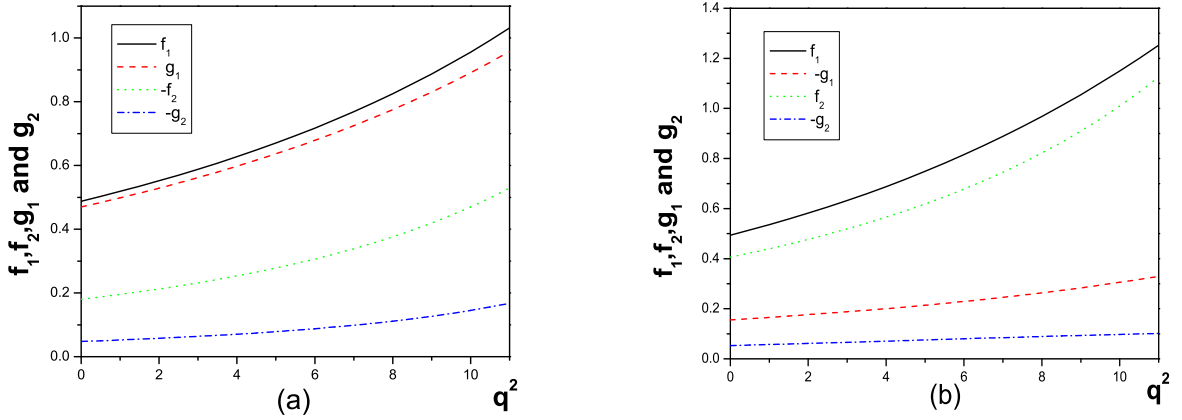}}
\end{center}
\caption{ Form factor for the decay $\Lambda_{b} \to \Lambda_{c}
l\bar{\nu}_l$ (a) and $\Sigma_{b} \to \Sigma_{c} l\bar{\nu}_l$
(b)}\label{f51}
\end{figure}

\begin{figure}[hhh]
\begin{center}
\scalebox{0.8}{\includegraphics{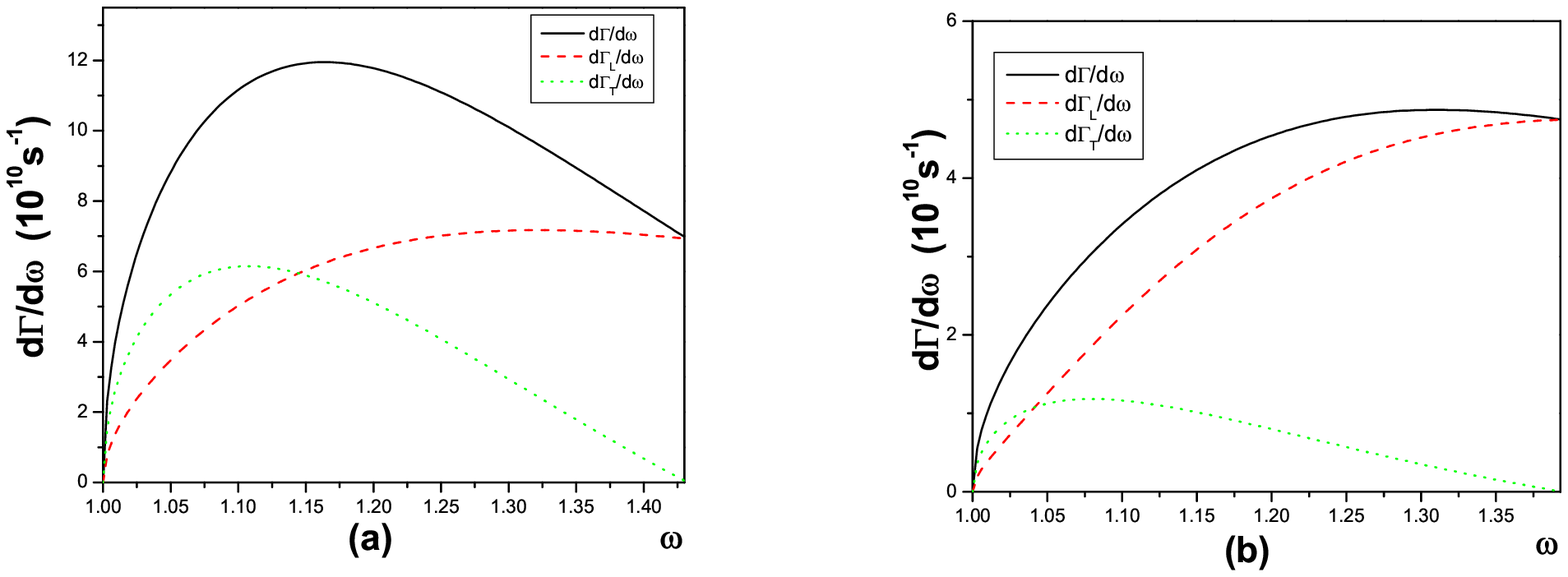}}
\end{center}
\caption{ Differential decay rates $d\Gamma/d\omega$ for the decay
$\Lambda_{b} \to \Lambda_{c} $ (a) and $\Sigma_{b} \to \Sigma_{c}
$ (b)}\label{f52}
\end{figure}

%\begin{figure}[hhh]
%\begin{center}
%\scalebox{0.8}{\includegraphics{dv.eps}}
%\end{center}
%\caption{Differential decay width for $\Xi_{cc} \rightarrow
%\Sigma_{c}e\nu_e$ at $m_{{[ud]}_V}=770$ MeV. }\label{f4}
%\end{figure}
Using the form factors, we evaluate the rates of $\Lambda_{b} \to
\Lambda_{c} l\bar{\nu}_l$. The decay rates are listed in table
\ref{Tab:t4}  where we
update  the results which were obtained in our earlier
paper\cite{Ke:2007tg} with the parameters fixed in
\cite{Chang:2018zjq} as inputs. Comparing them with that obtained by other
authors in
Ref.\cite{Ke:2007tg,Ebert:2006rp,Ivanov:1996fj,Ivanov:1998ya}, one
can notice that the differences among all the results are rather
small. We plot the differential decay rates $d\Gamma/d\omega$ in
Fig. \ref{f52}(a) which also is consistent with those given in
\cite{Ebert:2006rp}.

\begin{table}
\caption{The widths and polarization asymmetries of $\Lambda_b\to
\Lambda_c l\bar{\nu}_l$ .}\label{Tab:t4}
\begin{ruledtabular}
\begin{tabular}{c|ccccc}
    &  $\Gamma$ ($10^{10}{\rm s}^{-1}$) &  $a_L$  &  $a_T$   & $R$    &  $P_L$ \\\hline
  this work& 4.22&-0.962&-0.766&1.54&-0.885  \\\hline
  Our result in \cite{Ke:2007tg}
    &  5.15        &   -0.932 & -0.601 & 1.47 & -0.798 \\\hline
     the update of the results  in \cite{Ke:2007tg}\footnote{ We re-set the parameters $\beta_{b[ud]}=0.601$ GeV
     and $\beta_{c[ud]}=0.5375$ GeV while the other parameters are unchanged. A phenomenological
     factor\cite{Korner:1992uw} $\frac{Q^2_0}{Q^2_0+Q^2}$=$\frac{3.22}{3.22+0.25}=0.928$ is included in
     every wave functions to compensate the non-point effect for the diquark. The three parameters $F(0)$,
     $a$, $b$ for $f_1$ are, 0.580,\,0.207,\,0.101\,,  for $g_1$ are 0.567,\,0.161,\,0.122\,,  for $f_2$ are
 -0.129,\,0.716,\,0.177\, and for $g_2$ are -0.0187,\,0.186,\,0.197.}
    &  4.69        &   -0.952 & -0.654 & 1.66 & -0.841 \\\hline
  relativistic quark model(in \cite{Ebert:2006rp})
    &  5.64       &   -0.940 & -0.600 & 1.61  & -0.810\\\hline
   relativistic three-quark model\cite{Ivanov:1996fj}
    &5.39        &   - & - &1.60   &-\\\hline
    the Bethe-Salpeter approach\cite{Ivanov:1998ya}  & 6.09       &   - & - & - &-
\end{tabular}
\end{ruledtabular}
\end{table}

\begin{table}
\caption{The widths and up-down asymmetries of non-leptonic decays
$\Lambda_b\to\Lambda_c M$.}\label{Tab:t6}
\begin{ruledtabular}
\begin{tabular}{c|cc|cc}
  & \multicolumn{2}{c|}{Our result in this work~~~~~~}
  & \multicolumn{2}{c}{ Our result in \cite{Ke:2007tg}\footnote{ Since there exist a mistake in the expressions
  of $P_1$ and $D$ (Eq (60)) in \cite{Ke:2007tg} and we correct them in Eq. (C5) in this appendix.
  In addition a factor 2 was missing in the formula for the transition
$\frac{1}{2}\to\frac{1}{2}+V$ given in\cite{Cheng:1996cs} and we
have discussed this issue with the authors of
Ref.\cite{Cheng:1996cs}, and then they have carefully checked this
formula and agreed with us. Therefore the $\Gamma$ and $\alpha$ of
$\Lambda_b^0\to\Lambda_c^+ V$  in \cite{Ke:2007tg} are changed
slightly here }~~~~}
\\\hline
  & $\Gamma$ ($10^{10}{\rm s}^{-1}$)               & $\alpha$ & $\Gamma$ ($10^{10}{\rm s}^{-1}$) & $\alpha$          \\\hline
  $\Lambda_b^0\to\Lambda_c^+ \pi^-$  & $0.261$  & $-0.999$
                        & $0.307$  & $-1$      \\\hline
  $\Lambda_b^0\to\Lambda_c^+ \rho^-$ & $0.769$  & $-0.875$
                        & $0.848$  & $-0.883$  \\\hline
  $\Lambda_b^0\to\Lambda_c^+ K^-$    & $0.0209$  & $-0.999$
                        & $0.0247$  & $-1$      \\\hline
  $\Lambda_b^0\to\Lambda_c^+ K^{*-}$  & $0.0398$  & $-0.836$
                        & $0.0440$  & $-0.846$  \\\hline
  $\Lambda_b^0\to\Lambda_c^+ a_1^-$  & $0.758$  & $-0.710$
                        & $0.838$  & $-0.726$  \\\hline
  $\Lambda_b^0\to\Lambda_c^+ D_s^-$  & $0.927$ & $-0.974$
                        & $0.932$ & $-0.982$  \\\hline
  $\Lambda_b^0\to\Lambda_c^+ D^{*-}_s$& $1.403$ & $-0.327$
                        & $1.566$  & $-0.360$  \\\hline
  $\Lambda_b^0\to\Lambda_c^+ D^-$  & $0.0355$ & $-0.979$
                        & $0.0410$  & $-0.986$  \\\hline
  $\Lambda_b^0\to\Lambda_c^+ {D^*}^-$& $0.0630$& $-0.371$
                        & $0.0702$  & $-0.403$
\end{tabular}
\end{ruledtabular}
\end{table}

From the theoretical aspects, calculating the corresponding
quantities for the non-leptonic decays seems to be much more
complicated than for the semi-leptonic ones. The theoretical
framework adopted in this work is based on the factorization
assumption, namely the hadronic transition matrix element is
factorized into a product of two independent matrix elements of
currents,
\begin{eqnarray}\label{s0}
&& \la \Lambda_{c}(P',S_z')M \mid \mathcal{H} \mid \Lambda_{b}(P,S_z) \ra  \nonumber \\
 &=&\frac{G_FV_{bc}V^*_{qq'}}{\sqrt{2}}\la M \mid
\bar{q'} \gamma^{\mu} (1-\gamma_{5}) q \mid 0\ra\la
\Lambda_{c}(P',S_z') \mid \bar{c} \gamma^{\mu} (1-\gamma_{5}) b
\mid \Lambda_{b}(P,S_z) \ra,
\end{eqnarray}
where the term $\la M \mid \bar{q'} \gamma^{\mu} (1-\gamma_{5}) q
\mid 0\ra$ is determined by the decay constant and the transition
$\Lambda_{b}\rightarrow \Lambda_{c}$ is calculated in the previous
sections.
%For the weak decays of mesons, such factorization
%approach seems to work well. We would believe that the factorization
%approach also works for the baryon case, especially as the diquark
%picture is employed.
Since the decays $\Lambda_{b}^0\to\Lambda_{c}+M^-$ is the
so-called color-favored transition, the factorization should be a
good approximation. The study on these non-leptonic decays can
check how close to reality the obtained form factors for the
heavy bottomed baryons would be. In table \ref{Tab:t6} we present the
results of this work and previous papers together. One can notice that they are very close
to each other, and it means that the heavy-quark-light-diquark
picture is indeed a good approximation.

\subsection{$\Sigma_{b}\to \Sigma_c$  form factors and some decay modes }

\begin{table}
\caption{The $\Sigma_{b}\to \Sigma_c$ form factors given in the
  three-parameter form.}\label{Tab:t22}
\begin{ruledtabular}
\begin{tabular}{cccc}
  $F$    &  $F(0)$ &  $a$  &  $b$ \\\hline
  $f^v_1$  &   0.494     &  1.73    & 1.40   \\
$f^v_2$  &   0.407$^a$     &   1.03    & 0.830  \\
  $g^v_1$  &      -0.156    &     1.03&  0.355 \\
  $g^v_2$  &      -0.0529$^a$    &    1.58  &2.74
\end{tabular}
\end{ruledtabular}
\end{table}

\begin{table}
\caption{The widths (in unit $10^{10}{\rm s}^{-1}$) and
polarization asymmetries  of $\Sigma_b\to \Sigma_c
l\bar{\nu}_l$.}\label{Tab:t5}
\begin{ruledtabular}
\begin{tabular}{c|ccccc}
    &  $\Gamma$ &  $a_L$  &  $a_T$ & $R$    &  $P_L$ \\\hline
in this work
   &    $1.56$      & $0.726$ & $-0.267$& $4.70$ & $0.552$ \\\hline
  in \cite{Ke:2012wa}\footnote{The $F(0)$ of $f_2$ and $g_2$
   should have an additional factor $\frac{m_{\Sigma_b}}{m_{\Sigma_b}m_{\Sigma_c}}$
   in table II in \cite{Ke:2012wa} so the results have some apparent change list here .}
    & $1.42$      &   $0.676$ & $-0.765$ & $4.17$ & $0.397$ \\\hline
 spectator-quark model \cite{Singleton:1990ye}&   $4.3$       &   - & -&10.7  & -\\\hline
  relativistic quark model\cite{Ebert:2006rp} &  $1.44$        &  -  & - &5.89  & - \\\hline
the Bethe-Salpeter approach\cite{Ivanov:1998ya}&   $1.65$ & - &-&-
& -\\\hline
  relativistic three-quark model\cite{Ivanov:1996fj}&   $2.23$       &   - & - &5.76  & -
\end{tabular}
\end{ruledtabular}
\end{table}

\begin{table}
\caption{The widths (in unit $10^{10}{\rm s}^{-1}$) and up-down
asymmetries of non-leptonic decays $\Sigma_b\to\Sigma_c
M$.}\label{Tab:t7}
\begin{ruledtabular}
\begin{tabular}{c|cc|cc}
  & \multicolumn{2}{c|}{this work~~~~} & \multicolumn{2}{c}{in \cite{Ke:2012wa}\footnote{ Since the reasons present in
  the footnotes in table IV and VI the results change explicitly  }~~~~~~}
  \\\hline
  & $\Gamma$                & $\alpha$ & $\Gamma$ & $\alpha$          \\\hline
  $\Sigma_b^0\to\Sigma_c^+ \pi^-$  & $0.161$  & $0.574$
   & $0.140$  & $0.514$\\\hline
  $\Sigma_b^0\to\Sigma_c^+ \rho^-$  & $0.443$  & $0.586$
  & $0.392$  & $0.537$ \\\hline
  $\Sigma_b^0\to\Sigma_c^+ K^-$   & $0.0131$  & $0.568$
   & $0.0115$  & $0.510$ \\\hline
  $\Sigma_b^0\to\Sigma_c^+ K^{*-}$   & $0.0224$  & $0.589$
  & $0.0200$  & $0.544$ \\\hline
  $\Sigma_b^0\to\Sigma_c^+ a_1^-$  & $0.395$  & $0.603$
  & $0.358$  & $0.571$\\\hline
  $\Sigma_b^0\to\Sigma_c^+ D_s^-$  & $0.743$ & $0.460$
  & $0.727$ & $0.396$\\\hline
  $\Sigma_b^0\to\Sigma_c^+ D^{*-}_s$& $0.547$  & $0.662$
  & $0.510$ & $0.691$ \\\hline
  $\Sigma_b^0\to\Sigma_c^+ D^-$   & $0.0277$  & $0.472$ & $0.0266$ & $0.408$
 \\\hline
  $\Sigma_b^0\to\Sigma_c^+ {D^*}^-$& $0.0256$  & $0.653$
  & $0.0238$& $0.672$
\end{tabular}
\end{ruledtabular}
\end{table}

%\begin{figure}[hhh]
%\begin{center}
%\scalebox{0.8}{\includegraphics{dv.eps}}
%\end{center}
%\caption{Differential decay width for $\Xi_{cc} \rightarrow
%\Sigma_{c}e\nu_e$ at $m_{{[ud]}_V}=770$ MeV. }\label{f4}
%\end{figure}
Now we calculate the form factors for the transition
$\Sigma_b\to\Sigma_c$. Using the values set for $\Lambda_b$ and
$\Lambda_c$, we determine the parameters $\beta_1$ and $\beta_1'$.
Generally the parameters $\beta_{23}$ and $\beta_{23}'$ would be
different from those for $\Lambda_b$ and $\Lambda_c$ because the
total spin of the $ud$ subsystem in $\Sigma_{b(c)}$ is 1 but that
is 0 in $\Lambda_{b(c)}$. The situation is
similar to the spin configurations
of a pseudoscalar vs a vector. Even though there are no data available,
we boldly use the same parameters gained from  $\Lambda_b\to\Lambda_c$ to make
our predictions and the validity or approximation would be tested in the future
measurements.

The parameters of the form factor are listed in table
\ref{Tab:t22}. As discussed in the footnote of Table \ref{Tab:t22}
there should be an additional factor
$m_{\Sigma_b}/(m_{\Sigma_b}+m_{\Sigma_c})$ for the $F_0$ of $f_2$
and $g_2$ \cite{Ke:2012wa}. Taking into account of this factor the
results are close to those given in \cite{Ke:2012wa}. The
differential decay rates $d\Gamma/d\omega$ are depicted  in Fig.
\ref{f52}(b) whose line-shape is fully consistent with those for
$\Omega_b\to\Omega_c$ in Fig. 8 of Ref. \cite{Ebert:2006rp}. It is
understandable because the decay $\Omega_b\to\Omega_c$ and
$\Sigma_b\to\Sigma_c$ are similar under the flavor SU(3) symmetry.
Since $\Omega_b$ does not decay via strong interaction, the weak
decays are dominant. The results of semi-leptonic decay of
$\Sigma_b\to\Sigma_c$ can be found in table \ref{Tab:t5}. The
total width $\Gamma$ and the longitudinal asymmetry $a_L$ are
close to our previous result but not the transverse asymmetry
$a_T$. The longitudinal polarization asymmetry $P_L$ deviate from
our previous result by a certain extent.

The interpretation of the difference is not trivial. When one
constructs the vertex function, the first principle is to retain
the momentum conservation, then there are two schemes to be
selected, i.e. whether let the diquark polarization depend on the
total momentum of the baryon $P$ or the momentum of the diquark
$p_2$. In our earlier paper \cite{Ke:2017eqo}, we discussed the
two schemes for constructing the vertex function of
$\Sigma_{b(c)}$. Comparing the numerical results of the transition
rates of $\Sigma_b\to\Sigma_c^*$, we found that the values
calculated with the two schemes only slightly deviated from each
other. Then the conclusion might be that the two schemes are
actually equivalent. In our study on the transition
\cite{Ke:2012wa}, we only adopted the scheme where the diquark
polarization only depends on momentum $p_2$, but not $P$. One conjecture is that
the momentum dependence may lead to the small deviations of the
transverse asymmetry $a_T$ as is shown in this work. The
difference, even though is not large, may imply a distinction of
the momentum dependence, which needs to be further
investigated and we will do it  in our following works.

%The reason may come from the vertex function in
%\cite{Ke:2012wa} and we also discussed the topic in
%\cite{Ke:2017eqo}. The widthes and up-down symmetries for
%$\Sigma_b\to\Sigma_cM$ are close to each other between two
%pictures of baryon.

\section{Conclusions and discussions}

In terms of the extended light front quark model we explore the leptonic and
non-leptonic weak decays
of $\Lambda_b\to\Lambda_c$ and $\Sigma_b\to\Sigma_c$ with the
three-quark picture of baryon.

Based on our earlier works for studying hyperon and meson decays
in terms of the LFQM, this work has two purposes. The first one is
to make a further confirmation of two possible ways of determining
the momentum dependence of the diqaurk polarization during
constructing the vertex functions of baryons. In our earlier
work\cite{Ke:2012wa} , we suggested a possible momentum dependence
of the diqaurk polarization. Namely, we conjectured that the
polarization of the diquark should depend on its momentum $p_2$.
Later \cite{Ke:2017eqo}, we employed another possible scheme where
the diquark polarization depends on the total momentum of the
baryon $P$. All the two schemes respect the principle of momentum
conservation. The numerical results on the transition rates of
$\Sigma_b\to\Sigma_c^*$ obtained in the two schemes are very close
to each other. In this work, we use neither of the two schemes to
evaluate the transition rates of $\Sigma_{b}\to\Sigma_{c}$ in
LFQM, because we treat the two light quarks as free individuals
rather than demanding them to reside in a ``diquark". Then we
compare the results with that obtained in terms of the diquark
pictures. We notice that the rates obtained in the two approaches
are very close, but for longitudinal and transverse asymmetrie
$a_T$ for the semi-leptonic decay of $\Sigma_{b}\to\Sigma_{c}$
which are more sensitive to the approaches, an obvious deviation
appears, i.e the values of $a_T$ are different for the two cases.
Therefore, there may exist a more profound physics which leads to
the difference. In our coming work we will further study those
schemes, namely investigate their validity. Beside a theoretical
analysis, the experimental data would compose an acceptable
probe-stone. Therefore, we suggest the experimentalists to carry
out high accuracy measurements on  $a_L$ and $a_T$ of
$\Sigma_{b}\to\Sigma_{c}$ transitions which would a great help.

The second purpose of this work is to test the reasonability and
application of the diqaurk picture which is definitely a good
approximation at least at the leading order according to the acute
discussions on this topic.  Therefore, in this work, we use the
three-body picture of baryons, i.e. treat the two spectator light
quarks as individual on-mass-shell constituents with definite
masses and momenta. Indeed, the two light quarks compose a
color-anti-triplet and spin-0 (for $\Lambda_{b(c)}$ or spin-1 (for
$\Sigma_{b(c)}$) subsystem. In this picture, we do not priori
assume they are loosely bound into a physical composite: diquark.
Comparing the numerical results of the decay rates of of
$\Lambda_b\to\Lambda_c$ and $\Sigma_b\to\Sigma_c$ with that
obtained in terms of the diquark which is supposed to be a
point-like boson with a form factor, we find them to be very
close. The consistency implies that the diquark picture indeed is
valid, namely, in a heavy baryon ($\Lambda_{b(c)}$, or
$\Sigma_{b(c)}$ etc.) two light quarks may be bound into a
physical system and can behave as a point-like boson, especially
when it serves as a spectator during the transition between two
heavy baryons.

\section*{Acknowledgement}

We thank Prof. Hai-Yang Cheng for helpful discussions.
This work is supported by the National Natural Science Foundation
of China (NNSFC) under the contract No. 11375128 and 11675082.

\appendix

\section{The form factor:}
In this appendix  we work out the full expressions of these form
factors $f^{s}_i\,{(i=1,2)}$ by expanding the corresponding
traces.

It is straightforward to
calculate the four traces
 \begin{eqnarray}\label{a1a2}
&& \frac{1}{8P^+P'^+}{\rm Tr}[
(\bar{P}\!\!\!\!\!\slash+M_0)\gamma^+
  (\bar{P'}\!\!\!\!\!\slash+M_0')
  (p_1\!\!\!\!\!\slash'+m'_1)\gamma^{+}
  (p_1\!\!\!\!\!\slash+m_1)  ]\nonumber\\&&=-(p_1-x_1\bar P)\cdot(p'_1-x'_1\bar
  P')+(x_1M_0+m_1)(x'_1M'_0+m'_1)\nonumber\\&&=k_{1\perp}\cdot
  k'_{1\perp}+(x_1M_0+m_1)(x'_1M'_0+m'_1),\\&& \frac{1}{8P^+P'^+}{\rm Tr}[
(\bar{P}\!\!\!\!\!\slash+M_0)\gamma^+\gamma_5
  (\bar{P'}\!\!\!\!\!\slash+M_0')
  (p_1\!\!\!\!\!\slash'+m'_1)\gamma^{+}\gamma_5
  (p_1\!\!\!\!\!\slash+m_1)  ]\nonumber\\&&=(p_1-x_1\bar P)\cdot(p'_1-x'_1\bar
  P')+(x_1M_0+m_1)(x'_1M'_0+m'_1)\nonumber\\&&=-k_{1\perp}\cdot
  k'_{1\perp}+(x_1M_0+m_1)(x'_1M'_0+m'_1),\\&& \frac{1}{8P^+P'^+}{\rm Tr}[
(\bar{P}\!\!\!\!\!\slash+M_0)\sigma^{i+}
  (\bar{P'}\!\!\!\!\!\slash+M_0')
  (p_1\!\!\!\!\!\slash'+m'_1)\gamma^{+}
  (p_1\!\!\!\!\!\slash+m_1)  ]\nonumber\\&&=(x'_1M'_0+m'_1)(p^i_{\perp}-x_1\bar P^i_{\perp})-(x_1M_0+m_1)(p'^i_{\perp}-x'_1\bar
  P'^i_{\perp})\nonumber\\&&=(x'_1M'_0+m'_1)k^{i}_{1\perp}-(x_1M_0+m_1)k^{'i}_{1\perp},\\&& \frac{1}{8P^+P'^+}{\rm Tr}[
(\bar{P}\!\!\!\!\!\slash+M_0)\sigma^{i+}\gamma_5
  (\bar{P'}\!\!\!\!\!\slash+M_0')
  (p_1\!\!\!\!\!\slash'+m'_1)\gamma^{+}\gamma_5
  (p_1\!\!\!\!\!\slash+m_1)  ]\nonumber\\&&=(x'_1M'_0+m'_1)(p^i_{\perp}-x_1\bar P^i_{\perp})-(x_1M_0+m_1)(p'^i_{\perp}-x'_1\bar
  P'^i_{\perp})\nonumber\\&&=(x'_1M'_0+m'_1)k^{i}_{1\perp}+(x_1M_0+m_1)k^{'i}_{1\perp},
 \end{eqnarray}
where $\bar P^+=P^+$, $\bar P^{'+}=P^{'+}$, $\bar
 P^i_{\perp}=P^i_{\perp}$, $P^{'i}_{\perp}=P^{'i}_{\perp}$,
 $p^{+}_1=x_1P^+$, $p^{'+}_1=x_1P^{'+}$, $p^{i}_{1\perp}=x_1P^i_{\perp}+k^{i}_{1\perp}$,
 ${p'}_{1\perp}^{i}=x_1P^{'i}_{\perp}+k^{'i}_{1\perp}$, $p_1\cdot \bar
 P=e_1M_0$, ${p'}_1\cdot \bar
 P'={e'}_1{M'}_0$ and $(p_1-x_1\bar P)\cdot(p'_1-x'_1\bar
  P')=-k_{1\perp}\cdot
  k'_{1\perp}$ have been used.
 The four traces are the same as those in
Ref.\cite{Cheng:2003sm}

Then it is also simple to deduce
the others
 \begin{eqnarray}\label{a5}
&&{\rm
Tr}[(\bar{P'}\!\!\!\!\!\slash-M_0')\gamma_{5}(p_2\!\!\!\!\!\slash+m_2)(\bar{P}\!\!\!\!\!\slash+M_0)
 \gamma_{5}(p_3\!\!\!\!\!\slash-m_3)]\nonumber\\&&=\{{M_0'}{m_3}{p_2\cdot \bar{P}} +{M_0'}{m_2}{p_3\cdot \bar{P}} +{p_2\cdot \bar{P'}}{p_3\cdot \bar{P}} +
   {p_2\cdot \bar{P}}{p_3\cdot \bar{P'}} +{m_2}{m_3}{\bar{P}\cdot \bar{P'}}\nonumber\\&& +{M0}\left[{M_0'}\left({m_2}{m_3} +{p_2\cdot p_3} \right)  +
      {m_3}{p_2\cdot \bar{P'}} +{m_2}{p_3\cdot \bar{P'}} \right] -
   {p_2\cdot p_3}{\bar{P}\cdot \bar{P'}} \}\nonumber\\&&=4[{M_0}{M_0'}( {e_2'}{e_3} + {e_2}{e_3'} +
    {e_3}{m_2} + {e_3'}{m_2} +
    {e_2}{m_3}+ {e_2'}{m_3} +
   {m_2}{m_3}) \nonumber \\&&+ \frac{{M_0}{M_0'}
       \left( -2{e_1}{M_0} + {{M_0}}^2 + {{m_1}}^2 - {{m_2}}^2 - {{m_3}}^2 \right) }{2} +
    \frac{{m_2}{m_3}\left( {{M_0}}^2 + {{M_0'}}^2 + {{q_\perp}}^2 \right) }{2}  \nonumber\\&&+
    \frac{\left( 2{e_1}{M_0} - {{M_0}}^2 - {{m_1}}^2 + {{m_2}}^2 + {{m_3}}^2 \right)
       \left( {{M_0}}^2 + {{M_0'}}^2 + {{q_\perp}}^2 \right) }{4}],
 \end{eqnarray}
where these relations
$\bar{P}\cdot \bar{P'}=({{M_0}}^2 + {{M_0'}}^2 + {{q_\perp}}^2)/2
$, $p_2\cdot \bar{P}={p'}_2\cdot \bar{P}=e_2M_0$, $p_3\cdot
\bar{P}={p'}_3\cdot \bar{P}=e_3M_0$, $p_2\cdot
\bar{P'}={p'}_2\cdot \bar{P'}={e'}_2{M'}_0$,  $p_3\cdot
\bar{P'}={p'}_3\cdot \bar{P'}={e'}_3{M'}_0$ and $p_2\cdot
p_3=({M_0}^2 + {m_1}^2 - {m_2}^2 - {m_3}^2 - 2 M_0e_1)/2$ are
needed.

With these explicit expressions of
the these traces the detailed forms of $f^{s}_i\,{(i=1,2)}$ and
$g^{s}_i\,{(i=1,2)}$ can be obtained,
\begin{eqnarray}\label{a6}
f^s_1
 &=& \int\frac{ d x_2 d^2 k_{2\perp}}{2(2\pi)^3}\frac{ d x_3 d^2 k_{3\perp}}{2(2\pi)^3}
 \frac{k_{1\perp}\cdot
  k'_{1\perp}+(x_1M_0+m_1)(x'_1M'_0+m'_1)}{\sqrt{M_0^3(m_1+e_1)
 (m_2+e_2)(m_3+e_3)(m_1'+e_1')
 (m_2'+e_2')(m_3'+e_3')}}\nonumber \\
  &&\times\frac{\phi_{\Lambda_{c}}^*(x',k'_{\perp})
  \phi_{\Lambda_{b}}(x,k_{\perp})}{16\sqrt{x_1x'_1}}4[{M_0}{M_0'}( {e_2'}{e_3} + {e_2}{e_3'} +
    {e_3}{m_2} + {e_3'}{m_2} +
    {e_2}{m_3}+ {e_2'}{m_3} +
   {m_2}{m_3}) \nonumber \\&&+ \frac{{M_0}{M_0'}
       \left( -2{e_1}{M_0} + {{M_0}}^2 + {{m_1}}^2 - {{m_2}}^2 - {{m_3}}^2 \right) }{2} +
    \frac{{m_2}{m_3}\left( {{M_0}}^2 + {{M_0'}}^2 + {{q_\perp}}^2 \right) }{2}  \nonumber\\&&+
    \frac{\left( 2{e_1}{M_0} - {{M_0}}^2 - {{m_1}}^2 + {{m_2}}^2 + {{m_3}}^2 \right)
       \left( {{M_0}}^2 + {{M_0'}}^2 + {{q_\perp}}^2 \right) }{4}],\nonumber\\
\frac{f^s_2}{M_{\Lambda_{b}}}
 &=& \frac{1}{q_{\perp}^i}\int\frac{ d x_2 d^2 k_{2\perp}}{2(2\pi)^3}\frac{ d x_3 d^2 k_{3\perp}}{2(2\pi)^3}
 \frac{(x_1M_0+m_1)k^{'i}_{1\perp}-(x'_1M'_0+m'_1)k^{i}_{1\perp}}{\sqrt{M_0^3(m_1+e_1)
 (m_2+e_2)(m_3+e_3)(m_1'+e_1')
 (m_2'+e_2')(m_3'+e_3')}}\nonumber \\
  &&\times\frac{\phi_{\Lambda_{c}}^*(x',k'_{\perp})
  \phi_{\Lambda_{b}}(x,k_{\perp})}{16\sqrt{x_1x'_1}} 4[{M_0}{M_0'}( {e_2'}{e_3} + {e_2}{e_3'} +
    {e_3}{m_2} + {e_3'}{m_2} +
    {e_2}{m_3}+ {e_2'}{m_3} +
   {m_2}{m_3}) \nonumber \\&&+ \frac{{M_0}{M_0'}
       \left( -2{e_1}{M_0} + {{M_0}}^2 + {{m_1}}^2 - {{m_2}}^2 - {{m_3}}^2 \right) }{2} +
    \frac{{m_2}{m_3}\left( {{M_0}}^2 + {{M_0'}}^2 + {{q_\perp}}^2 \right) }{2}  \nonumber\\&&+
    \frac{\left( 2{e_1}{M_0} - {{M_0}}^2 - {{m_1}}^2 + {{m_2}}^2 + {{m_3}}^2 \right)
       \left( {{M_0}}^2 + {{M_0'}}^2 + {{q_\perp}}^2 \right) }{4}],
\nonumber\\
g^s_1
 &=& \int\frac{ d x_2 d^2 k_{2\perp}}{2(2\pi)^3}\frac{ d x_3 d^2 k_{3\perp}}{2(2\pi)^3}
 \frac{-k_{1\perp}\cdot
  k'_{1\perp}+(x_1M_0+m_1)(x'_1M'_0+m'_1)}{\sqrt{M_0^3(m_1+e_1)
 (m_2+e_2)(m_3+e_3)(m_1'+e_1')
 (m_2'+e_2')(m_3'+e_3')}}\nonumber \\
  &&\times\frac{\phi_{\Lambda_{c}}^*(x',k'_{\perp})
  \phi_{\Lambda_{b}}(x,k_{\perp})}{16\sqrt{x_1x'_1}} 4[{M_0}{M_0'}( {e_2'}{e_3} + {e_2}{e_3'} +
    {e_3}{m_2} + {e_3'}{m_2} +
    {e_2}{m_3}+ {e_2'}{m_3} +
   {m_2}{m_3}) \nonumber \\&&+ \frac{{M_0}{M_0'}
       \left( -2{e_1}{M_0} + {{M_0}}^2 + {{m_1}}^2 - {{m_2}}^2 - {{m_3}}^2 \right) }{2} +
    \frac{{m_2}{m_3}\left( {{M_0}}^2 + {{M_0'}}^2 + {{q_\perp}}^2 \right) }{2}  \nonumber\\&&+
    \frac{\left( 2{e_1}{M_0} - {{M_0}}^2 - {{m_1}}^2 + {{m_2}}^2 + {{m_3}}^2 \right)
       \left( {{M_0}}^2 + {{M_0'}}^2 + {{q_\perp}}^2 \right) }{4}],
\nonumber\\
\frac{g^s_2}{M_{\Lambda_{b}}}
 &=& \frac{1}{q_{\perp}^i}\int\frac{ d x_2 d^2 k_{2\perp}}{2(2\pi)^3}\frac{ d x_3 d^2 k_{3\perp}}{2(2\pi)^3}
 \frac{(x_1M_0+m_1)k^{'i}_{1\perp}+(x'_1M'_0+m'_1)k^{i}_{1\perp}}{\sqrt{M_0^3(m_1+e_1)
 (m_2+e_2)(m_3+e_3)(m_1'+e_1')
 (m_2'+e_2')(m_3'+e_3')}}\nonumber \\
  &&\times\frac{\phi_{\Lambda_{c}}^*(x',k'_{\perp})
  \phi_{\Lambda_{b}}(x,k_{\perp})}{16\sqrt{x_1x'_1}} 4[{M_0}{M_0'}( {e_2'}{e_3} + {e_2}{e_3'} +
    {e_3}{m_2} + {e_3'}{m_2} +
    {e_2}{m_3}+ {e_2'}{m_3} +
   {m_2}{m_3}) \nonumber \\&&+ \frac{{M_0}{M_0'}
       \left( -2{e_1}{M_0} + {{M_0}}^2 + {{m_1}}^2 - {{m_2}}^2 - {{m_3}}^2 \right) }{2} +
    \frac{{m_2}{m_3}\left( {{M_0}}^2 + {{M_0'}}^2 + {{q_\perp}}^2 \right) }{2}  \nonumber\\&&+
    \frac{\left( 2{e_1}{M_0} - {{M_0}}^2 - {{m_1}}^2 + {{m_2}}^2 + {{m_3}}^2 \right)
       \left( {{M_0}}^2 + {{M_0'}}^2 + {{q_\perp}}^2 \right) }{4}].
\end{eqnarray}

By the same way the traces in
the form factors  $f^{v}_i\,{(i=1,2)}$ and $g^{v}_i\,{(i=1,2)}$
also can be directly calculated. Since they are very long, we omit them for saving space.
Since these form factors $f^{s(v)}_i\,{(i=1,2)}$ and
$g^{s(v)}_i\,{(i=1,2)}$ are evaluated in the frame $q^+=0$ i.e.
$q^2=-q^2_{\perp}\leq 0$ (the space-like region) one needs to
extend them into the time-like region. One can employ a
three-parameter form\cite{pentaquark2}
 \begin{eqnarray}\label{s145}
 F(q^2)=\frac{F(0)}{\left(1-\frac{q^2}{M_{\mathcal{B}_{b}}^2}\right)
  \left[1-a\left(\frac{q^2}{M_{\mathcal{B}_{b}}^2}\right)
  +b\left(\frac{q^2}{M_{\mathcal{B}_b}^2}\right)^2\right]},
 \end{eqnarray}
where $F(q^2)$ denotes the form factors $f^{s(v)}_i$ and
$g^{s(v)}_i$.
 Using the form factors in the space-like region
we may numerically calculate  the parameters $a,~b$ and $F(0)$ in
the un-physical region, namely fix $F(q^2\leq 0)$. As discussed in
previous section, these forms are extended into the physical
region with $q^2\geq 0$ through Eq.(\ref{s145}).

\section{Semi-leptonic decays of  $\mathcal{B}_b\to
\mathcal{B}_c  l\bar\nu_l$ }

The helicity amplitudes are related to the form factors for
$\mathcal{B}_b\to \mathcal{B}_c l\bar\nu_l$ through the following
expressions \cite{Korner:1991ph,Bialas:1992ny,Korner:1994nh}
 \beq
 H^V_{\frac{1}{2},0}&=&\frac{\sqrt{Q_-}}{\sqrt{q^2}}\left(
  \left(\Mb+\Mc\right)f_1-\frac{q^2}{\Mb}f_2\right),\non\\
 H^V_{\frac{1}{2},1}&=&\sqrt{2Q_-}\left(-f_1+
  \frac{\Mb+\Mc}{\Mb}f_2\right),\non\\
 H^A_{\frac{1}{2},0}&=&\frac{\sqrt{Q_+}}{\sqrt{q^2}}\left(
  \left(\Mb-\Mc\right)g_1+\frac{q^2}{\Mb}g_2\right),\non\\
 H^A_{\frac{1}{2},1}&=&\sqrt{2Q_+}\left(-g_1-
  \frac{\Mb-\Mc}{\Mb}g_2\right).
 \eeq
where $Q_{\pm}=2(P\cdot P'\pm \Mb\Mc)$ and $\Mb\, (\Mc)$
represents $M_{\mathcal{B}_b}$ or  $M_{\mathcal{B}_c}$. The
amplitudes for the negative helicities are obtained in terms of
the relation
 \beq
 H^{V,A}_{-\lambda'-\lambda_W}=\pm H^{V,A}_{\lambda',\lambda_W},
  \eeq
where the upper (lower) sign corresponds to V(A).
 The helicity
amplitudes are
 \beq
 H_{\lambda',\lambda_W}=H^V_{\lambda',\lambda_W}-
  H^A_{\lambda',\lambda_W}.
 \eeq
The helicities of the $W$-boson $\lambda_W$ can be either $0$ or
$1$, which correspond to the longitudinal and transverse
polarizations, respectively.  The longitudinally (L) and
transversely (T) polarized rates are
respectively\cite{Korner:1991ph,Bialas:1992ny,Korner:1994nh}
 \beq
 \frac{d\Gamma_L}{d\omega}&=&\frac{G_F^2|V_{cb}|^2}{(2\pi)^3}~
  \frac{q^2~p_c~\Mc}{12\Mb}\left[
  |H_{\frac{1}{2},0}|^2+|H_{-\frac{1}{2},0}|^2\right],\non\\
 \frac{d\Gamma_T}{d\omega}&=&\frac{G_F^2|V_{cb}|^2}{(2\pi)^3}~
  \frac{q^2~p_c~\Mc}{12\Mb}\left[
  |H_{\frac{1}{2},1}|^2+|H_{-\frac{1}{2},-1}|^2\right].
 \eeq
where $p_c$ is the momentum of $\mathcal{B}_c$ in the reset frame
of $\mathcal{B}_b$.

The integrated longitudinal and transverse asymmetries defined as
 \beq
 a_L&=&\frac{\int_1^{\omega_{\rm max}} d\omega ~q^2~ p_c
     \left[ |H_{\frac{1}{2},0}|^2-|H_{-\frac{1}{2},0}|^2\right]}
     {\int_1^{\omega_{\rm max}} d\omega ~q^2~ p_c
     \left[|H_{\frac{1}{2},0}|^2+|H_{-\frac{1}{2},0}|^2\right]},
     \non\\
 a_T&=&\frac{\int_1^{\omega_{\rm max}} d\omega ~q^2~ p_c
     \left[ |H_{\frac{1}{2},1}|^2-|H_{-\frac{1}{2},-1}|^2\right]}
     {\int_1^{\omega_{\rm max}} d\omega ~q^2~ p_c
     \left[|H_{\frac{1}{2},1}|^2+|H_{-\frac{1}{2},-1}|^2\right]}.
 \eeq
 The ratio of the longitudinal to
transverse decay rates $R$ is defined by
 \beq
 R=\frac{\Gamma_L}{\Gamma_T}=\frac{\int_1^{\omega_{\rm
     max}}d\omega~q^2~p_c\left[ |H_{\frac{1}{2},0}|^2+|H_{-\frac{1}{2},0}|^2
     \right]}{\int_1^{\omega_{\rm max}}d\omega~q^2~p_c
     \left[ |H_{\frac{1}{2},1}|^2+|H_{-\frac{1}{2},-1}|^2\right]},
 \eeq
and the  longitudinal polarization asymmetry $P_L$ is given as
 \beq
 P_L&=&\frac{\int_1^{\omega_{\rm max}} d\omega ~q^2~ p_c
     \left[ |H_{\frac{1}{2},0}|^2-|H_{-\frac{1}{2},0}|^2+
     |H_{\frac{1}{2},1}|^2-|H_{-\frac{1}{2},-1}|^2\right]}
     {\int_1^{\omega_{\rm max}} d\omega ~q^2~ p_c
     \left[|H_{\frac{1}{2},0}|^2+|H_{-\frac{1}{2},0}|^2+
     |H_{\frac{1}{2},1}|^2+|H_{-\frac{1}{2},-1}|^2\right]}.
 \eeq
\section{$\mathcal{B}_b\to
\mathcal{B}_c M$} In general, the transition amplitude of
$\mathcal{B}_b\to \mathcal{B}_c M$ can be written as
 \beq
 {\cal M}(\mathcal{B}_b\to
\mathcal{B}_c P)&=&\bar
  u_{\Lambda_c}(A+B\gamma_5)u_{\Lambda_b}, \non \\
 {\cal M}(\mathcal{B}_b\to
\mathcal{B}_c V)&=&\bar
  u_{\Lambda_c}\epsilon^{*\mu}\left[A_1\gamma_{\mu}\gamma_5+
   A_2(p_c)_{\mu}\gamma_5+B_1\gamma_{\mu}+
   B_2(p_c)_{\mu}\right]u_{\Lambda_b},
 \eeq
where $\epsilon^{\mu}$ is the polarization vector of the final
vector or axial-vector mesons. Including the effective Wilson
coefficient $a_1=c_1+c_2/N_c$, the decay amplitudes in the
factorization approximation are \cite{Korner:1992wi,Cheng:1996cs}
 \beq
 A&=&\lambda f_P(\Mb-\Mc)f_1(M^2), \non \\
 B&=&\lambda f_P(\Mb+\Mc)g_1(M^2), \non\\
 A_1&=&-\lambda f_VM\left[g_1(M^2)+g_2(M^2)\frac{\Mb-\Mc}{\Mb}\right],
 \non\\
 A_2&=&-2\lambda f_VM\frac{g_2(M^2)}{\Mb},\non\\
 B_1&=&\lambda f_VM\left[f_1(M^2)-f_2(M^2)\frac{\Mb+\Mc}{\Mb}\right],
 \non\\
 B_2&=&2\lambda f_VM\frac{f_2(M^2)}{\Mb},
 \eeq
where $\lambda=\frac{G_F}{\sqrt 2}V_{cb}V_{q_1q_2}^*a_1$ and $M$
is the meson mass. Replacing  $P$, $V$ by $S$ and $A$ in the above
expressions, one can easily obtain similar expressions for scalar
and axial-vector mesons .

The decay rates of $\mathcal{B}_b\rightarrow \mathcal{B}_cP(S)$
and up-down asymmetries are \cite{Cheng:1996cs}
 \begin{eqnarray}
 \Gamma&=&\frac{p_c}{8\pi}\left[\frac{(\Mb+\Mc)^2-M^2}{\Mb^2}|A|^2+
  \frac{(\Mb-\Mc)^2-M^2}{\Mb^2}|B|^2\right], \non\\
 \alpha&=&-\frac{2\kappa{\rm Re}(A^*B)}{|A|^2+\kappa^2|B|^2},
 \end{eqnarray}
where $p_c$ is the $\mathcal{B}_c$ momentum in the rest frame of
$\mathcal{B}_b$ and $\kappa=\frac{p_c}{E_{\mathcal{B}_c}+\Mc}$.
For $\mathcal{B}_b\rightarrow \mathcal{B} V(A)$ decays, the decay
rates and up-down asymmetries are
 \beq
 \Gamma&=&\frac{p_c (E_{\Lambda_c}+\Mc)}{4\pi\Mb}\left[
  2\left(|S|^2+|P_2|^2\right)+\frac{E^2}{M^2}\left(
  |S+D|^2+|P_1|^2\right)\right], \non\\
 \alpha&=&\frac{4M^2{\rm Re}(S^*P_2)+2E^2{\rm Re}(S+D)^*P_1}
  {2M^2\left(|S|^2+|P_2|^2 \right)+E^2\left(|S+D|^2+|P_1|^2
  \right) },
 \eeq
where $E$ is energy of the vector (axial vector) meson, and
 \begin{eqnarray}
  S&=&-A_1, \non\\
  P_1&=&-\frac{p_c}{E}\left(\frac{\Mb+\Mc}
  {E_{\Lambda_c}+\Mc}B_1+M_bB_2\right), \non \\
  P_2&=&\frac{p_c}{E_{\Lambda_c}+\Mc}B_1,\non\\
  D&=&-\frac{p^2_c}{E(E_{\Lambda_c}+\Mc)}(A_1- M_bA_2).
 \end{eqnarray}

\end{document}